\documentclass[aps,11pt,showpacs,superscriptaddress,amsfonts,amssymb,amsmath]{revtex4}

\usepackage{graphicx}

\begin{document}

\title{Comparison of metrics from retarded integrals and transverse traceless subgauge}

\author{R.A.\ Lewis \footnote{Email address: r3l@psu.edu}}
 \affiliation{Penn State University (ret.), Boalsburg, PA , USA}
\author{G.\ Modanese \footnote{Email address: giovanni.modanese@unibz.it}}
 \affiliation{Free University of Bolzano, Faculty of Science and Technology, Bolzano, Italy}

\date{Apr 5, 2013}

\begin{abstract}

The time-varying gravitational field produced by a Weber bar is used to explore mathematical features of the linearized Einstein equation. We present a self-contained formal framework for the treatment of the linear field, which is applicable to several situations where the standard quadrupolar formulas are not adequate. The expressions for retarded integrals reveal a singularity associated with boundary conditions.  Results from the transverse traceless subgauge are compared with the radiation calculated from retarded integrals.  Lienard-Wiechert potentials are used in a treatment of the Weber bar as a collection of point particles and further possible applications are outlined.  The Riemann tensor clarifies the transition from near-field geodesic forces to tidal forces in the far field. 

\end{abstract}


 \maketitle

\section{Introduction} 

In this paper, the Weber bar is used as a specific example of a gravitational source, to illustrate various aspects of the mathematics and physics of the linearized Einstein equation.  The discussion comprises techniques for computing retarded integrals, including Lienard-Wiechert potentials.  These techniques are complementary to less detailed discussions of more general sources found in textbooks.  For example, boundary conditions are a crucial aspect of the Weber bar, in that the outer few grams of material contribute as much to the gravitational disturbance as the inner hundreds of kilograms of mass.  The Riemann tensor interpolates smoothly between gravitoelectric fields in the near field and tidal forces in the far radiation field. 

The aim of this work is to present a self-contained formal framework for the computation of the linearized field in situations which are different from those considered in standard textbooks. The typical approximation methods employed in the literature, based on the oscillating-quadrupole formula, involve several simplifying assumptions: far fields (radiation region), no retardation in the source, symmetry of plane wave or spherical wave, TT gauge and absence of longitudinal field components. We relax all these assumptions. We also avoid the use of power series and Fourier expansions in the source, which might fail if sharp discontinuities are present. We provide an example of numerical and analytical computation of the near field. Within our framework, we describe the effects of the fields on detectors through standard formal ingredients (Riemann tensor, geodesic deviation, gravito-electric field components), but also considering possible forces and strains which are not typical of plane transverse waves.

 In Sect.\ \ref{grav} the metric produced by a Weber bar is related to retarded integrals of the stress tensor.  A mixture of Lagrange and Euler variables is used to establish notation.  The technique for treating boundary conditions is shown explicitly.  Section \ref{tran} describes the procedure used to express the quadrupole component of the radiation in the transverse traceless (TT) subgauge.  Section \ref{gaug} discusses the (non) existence of a global gauge transformation between the harmonic gauge and the TT gauge. A local transformation can always be obtained, however.

    Section \ref{det} describes expressions used to calculate force gradients in a detector.  Gradients in the gravitoelectric field and the second time derivative of the spatial components of the metric are treated as Newtonian forces.  The Riemann tensor, which includes both geodesic and tidal forces, is introduced. In Sect.s \ref{x-axis} and \ref{z-axis} the response of a detector located on coordinate axes is expressed, resulting in the same response using either the retarded integral or transverse traceless metrics.  

In section \ref{near}, expressions for near field components of the metric are evaluated, taking into account higher order terms in the retarded integral.  Time-dependent metric components varying as $1/r^2$ and $1/r^3$ are computed.  A numerical integration shows how metric components evolve from near-field to radiation field phenomena.  

    In section \ref{off}, the detector is located in a direction not on a coordinate axis.  The descriptions of the forces in terms of the Riemann tensor are the same for the retarded integral and TT metrics, even though different components of the metrics contribute to Riemann tensor elements.  The agreement between retarded integrals and the TT metric is important, since the derivation of this relationship in some texts ignores boundary conditions.  

     In section \ref{time}, the Weber bar is treated as a collection of point particles, using Lienard-Wiechert potentials to express the gravitational field.   The result agrees with that computed using density as a continuum distribution. In section  \ref{app} we present in a qualitative way some possible future applications of the method of Sect.\ 
 \ref{time}, for instance to the computation of gravitational fields generated by stationary waves in a plasma.

\section{Gravitational field of an axial oscillator}
\label{grav}
 
The axial oscillator considered in this paper is similar to a Weber bar, which oscillates longitudinally in the $z$ direction (Fig.\ 1).  The linearized Einstein equation relates trace-reversed metric deviation elements and energy-momentum tensor elements as follows 
\begin{equation}
{\nabla ^2}{\bar h_{\mu \nu }} - \frac{{{\partial ^2}{{\bar h}_{\mu \nu }}}}{{{c^2}\partial {t^2}}}   =  - 16\pi G{T_{\mu \nu }}
\label{eq1}
\end{equation}
Metric deviations are expressed in terms of the metric as follows
\begin{equation}
{g_{\mu \nu }}(x) = {\eta _{\mu \nu }} + {h_{\mu \nu }}(x); \ \ \ \ 
{\eta _{\mu \nu }} = {\rm{diag}}\left( { - 1,1,1,1} \right)
\end{equation}
The trace-reversed metric deviation is defined as
\begin{equation}
{\bar h_{\mu \nu }}(x) = {h_{\mu \nu }}(x) - \frac{1}{2}{\eta _{\mu \nu }}h_\lambda ^\lambda (x)
\end{equation}
and satisfies the harmonic gauge condition (analogous to the Lorentz gauge in electrodynamics)
\begin{equation}
\partial^\mu {\bar h_{\mu \nu }}(x) = 0
\end{equation}
The solution to the Einstein equation can be expressed as a retarded integral
\begin{equation}
{\bar h^{\mu \nu }}\left( {{{\bf x}_0},{t_0}} \right) = \frac{{4G}}{{{c^4}}}\int {dt} \int {{d^3}} x\frac{{{T^{\mu \nu }}\left( {\bf x},t \right)}}{{\left| {{{\bf x}_0} - \bf x} \right|}}\delta \left( {{t_0} - t + \frac{{\left| {{{\bf x}_0} - \bf x} \right|}}{c}} \right)
\label{general-ret-integral}
\end{equation}
For points outside the source, the fields (\ref{general-ret-integral}) can be evaluated using the series expansions
 \begin{equation}
\frac{1}{{\left| {{{\bf x}_0} -{\bf x}} \right|}} = {\left( {{\bf x}_0^2 - 2{{\bf x}_0} \cdot {\bf x} + {{\bf x}^2}} \right)^{ - \frac{1}{2}}} = \frac{1}{{{r_0}}} + \frac{{{\bf x} \cdot {\bf n}}}{{r_0^2}} + \frac{{3{{\left( {{\bf x} \cdot {\bf n}} \right)}^2} - {{\bf x}^2}}}{{2r_0^3}} + ...
\label{series1}
\end{equation}
\begin{equation}
\left| {{{\bf x}_0} - \bf x} \right| = {\left( {\bf x_0^2 - 2{{\bf x}_0} \cdot \bf x + {{\bf x}^2}} \right)^{\frac{1}{2}}} = {r_0} - {\bf n} \cdot {\bf x} + \frac{{{{\bf x}^2} - {{\left( {{\bf n} \cdot {\bf x}} \right)}^2}}}{{2{r_0}}} + ...
\label{series2}
\end{equation}
It is convenient to express the metric deviation as a sum of near and far fields.  Time-dependent terms of first order in $1/r_0$  correspond to radiation (energy transported to infinity), while terms of higher order refer to near fields.  

The energy-momentum tensor can be written in two different ways. In ``Lagrangian'' variables \cite{Kittel} one has
\begin{equation}
{T^{\mu \nu }}_{Lagrange} = {\left( {\begin{array}{*{20}{c}}
{\rho {c^2}}&0&0&{\rho c\frac{{\partial w}}{{\partial t}}}\\
0&0&0&0\\
0&0&0&0\\
{\rho c\frac{{\partial w}}{{\partial t}}}&0&0&{ - E\frac{{\partial w}}{{\partial z}}}
\end{array}} \right)}
\label{T-Lagr}
\end{equation}
where $\rho$ is the mass density, $w$ a longitudinal strain variable and $E$ the elastic modulus of the material. Alternatively, one can use ``Euler'' variables \cite{Weinberg}
\begin{equation}
{T^{\mu \nu }}_{Euler} =
{\left( {\begin{array}{*{20}{c}}
{\rho {c^2}}&0&0&{\rho vc}\\
0&0&0&0\\
0&0&0&0\\
{\rho vc}&0&0&{{\rho _1}v_s^2}
\end{array}} \right)}
\label{T-Eul}
\end{equation}
where $\rho_1$ is a density perturbation ($\rho=\rho_0+\rho_1$), $v$ is the material velocity and $v_s$ is the speed of sound in the material, given by ${v_s} = \sqrt {\frac{E}{{{\rho _0}}}}$. For an ideal elastic material, there are known continuity and stress-strain relations connecting $\frac{{\partial w}}{{\partial z}}$, $\frac{{\partial w}}{{\partial t}}$ and $v$ at any point. The component $T^{03}$ of the energy-momentum tensor is equal to the mass current $J_z$ multiplied by $c$, and the component $T^{33}$ is also called the pressure $p$.

\begin{figure}
\begin{center}
  \includegraphics[width=10cm,height=7cm]{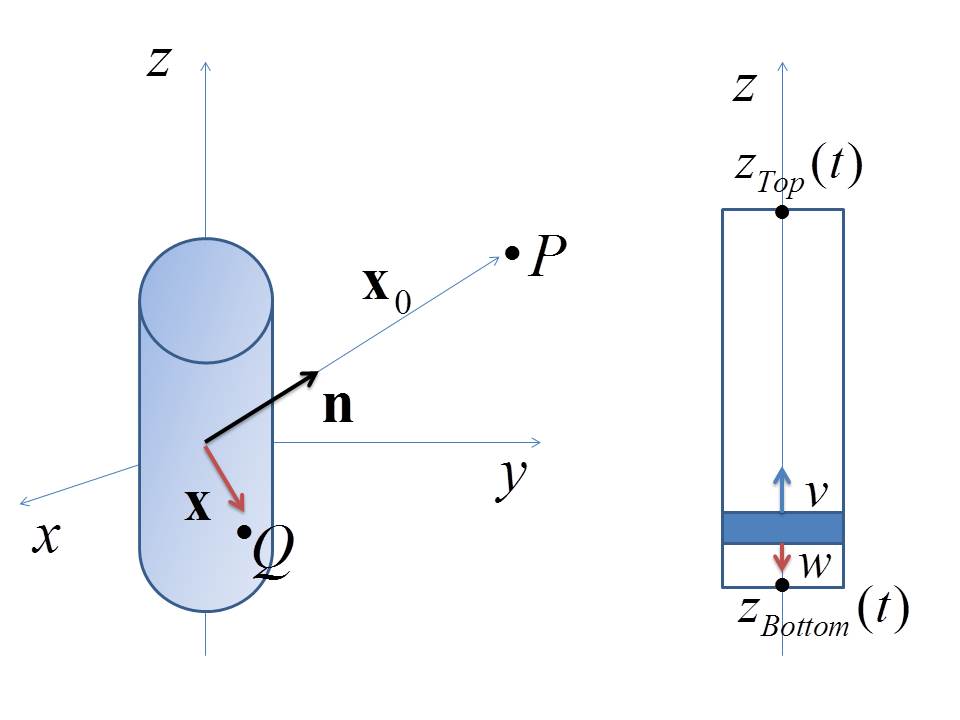}
\caption{Weber bar with longitudinal oscillations in the $z$ direction. A generic point $Q$ of the bar, with coordinate ${\bf x}$, has at a given instant displacement $w$ with respect to its equilibrium position, and velocity $v$. Both $w$ and $v$ are directed along $z$, and can be positive or negative. The retarded integral (\ref{general-ret-integral}) gives the metric at an observation point $P$ with coordinate ${\bf x_0}$. The upper and lower boundaries of the bar have coordinates $z_{Top}$ and $z_{Bottom}$ which vary in time.} 
\end{center}      
\end{figure}

When the bar oscillates in its lowest frequency normal mode, the longitudinal displacement of any point of the bar with respect to its equilibrium position can be expressed as follows
\begin{equation}
w(z,t) = \varepsilon L\sin \left( {\frac{{\pi z}}{L}} \right)\cos \omega t ,
\ \ \ \ \omega  = \frac{\pi }{L} v_s
\end{equation}
where $L$ is the length of the bar,  $\varepsilon \ll 1$ is a perturbation parameter, and $\omega$ is the resonant frequency.

Expressions for mass current and  the pressure are straightforward.  The pressure is written as 
\begin{equation}
p\left( {z,t} \right) = {\rho _1}v_s^2 =  - E\frac{{\partial w}}{{\partial z}} =  - {\omega ^2}{L^2}{\rho _0}\frac{\varepsilon }{\pi }\cos \left( {\frac{{\pi z}}{L}} \right)\cos \omega t
\end{equation}
The current density is expressed as follows
\begin{equation}
{J_z}\left( {z,t} \right) = \rho \frac{{\partial w}}{{\partial t}} =  - \varepsilon {\rho _0}\omega L\sin \left( {\frac{{\pi z}}{L}} \right)\sin \omega t
\end{equation}

In this section, only radiation is described, since the mathematics is simpler.  Evaluation of near fields (which is impossible in the TT gauge) is postponed to Sect.\ \ref{near}.  As usual in the calculation of the radiation field, the denominator $|{{\bf x}_0} - \bf x|$  in eq.\ (\ref{general-ret-integral}) is replaced by $r_0=|\bf x_0|$ (distance from the center of the Weber bar to the observation point $P$). This is justified because the wavelength $\lambda  = \frac{{2\pi \omega }}{c}$  of the radiation is very long compared with the dimensions of the bar. In the following, we shall write explicitly this kind of approximation only when needed.

The retarded integrals of the pressure and of the mass current are quite straightforward, because these quantities are of first order in $\varepsilon$, so it is not necessary for their computation to take into account the movement of the upper and lower boundaries of the bar. That movement is itself of order $\varepsilon$, and plays a role only in the computation of the integral of $T^{00}$ (see below).
The retarded integral of the pressure is as follows 
\begin{equation}
{\bar h^{33}}\left( {{{\bf{x}}_0},{t_0}} \right) = \frac{{4G}}{{{c^4}}}\int {{d^3}x} \, p\left( {{\bf{x}},{t_0} - \frac{{{r_0}}}{c}} \right) =  - \frac{{8G}}{{{r_0}{c^4}}}M{L^2}{\omega ^2}\frac{\varepsilon }{{{\pi ^2}}}\cos \left( {\omega {t_0} - k{r_0}} \right)
\end{equation}
where $M$ is the mass of the Weber bar.  Evaluating the retarded integral of the mass current requires a position-dependent expression for the retarded time,
\begin{equation}
{\bar h^{03}}\left( {{{\bf x}_0},{t_0}} \right) = \frac{{4G}}{{{r_0}{c^3}}}\int {{d^3}x} \, J\left( {{\bf{x}},{t_0} - \frac{{{r_0} - {\bf{n}} \cdot {\bf{x}}}}{c}} \right) =  - \frac{{8G}}{{{r_0}{c^4}}}M{L^2}{\omega ^2}{n_z}\frac{\varepsilon }{{{\pi ^2}}}\cos \left( {\omega {t_0} - k{r_0}} \right)
\label{eq65}
\end{equation}
where ${\bf n}=(n_x,n_y,n_z)$  is a unit vector pointing towards the observation point.

The expression for the retarded current contribution involves the integral 
\begin{equation}
\begin{array}{l}
{I^{03}} = \int\limits_{ - L/2}^{L/2} {dz} \sin \left( {\frac{{\pi z}}{L}} \right)\sin \left( {\omega {t_0} - k{r_0} + k{n_z}z} \right)\\
 \cong \int\limits_{ - L/2}^{L/2} {dz} \sin \left( {\frac{{\pi z}}{L}} \right)\left[ {\sin \left( {\omega {t_0} - k{r_0}} \right) + k{n_z}z\cos \left( {\omega {t_0} - k{r_0}} \right)} \right] = 2\frac{{{L^2}}}{{{\pi ^2}}}k{n_z}\cos \left( {\omega {t_0} - k{r_0}} \right)
\end{array}
\end{equation}
using the long wavelength approximation $kL \ll 1$.  

      Evaluating the retarded integral of the mass density $T^{00}$ is more elaborate.  The density is written as a sum of an initial constant density $\rho_0$ and perturbed $\rho_1$ terms, where 
\begin{equation}
{\rho _1}\left( {z,t} \right) =  - \varepsilon {\rho _0}\pi \cos \left( {\frac{{\pi z}}{L}} \right)\cos \left( {\omega t} \right)
\label{eq14}
\end{equation}

The integral for $\bar h^{00}$  has a hidden singularity, which is revealed by expressing only the time-varying component 
\begin{equation}
{\bar h^{00}}\left( {{{\bf{x}}_0},{t_0}} \right) = \frac{{4GA}}{{{r_0}{c^4}}}\left[ {\int\limits_{{z_{Bottom}}(t_0)}^{{z_{Top}}\left( t_0 \right)} {dz\rho \left( {{\bf{x}},{t_0} - \frac{{\left| {{{\bf{x}}_0} - {\bf{x}}} \right|}}{c}} \right) - \int\limits_{ - L/2}^{L/2} {dz} {\rho _0}} } \right]
\end{equation}
where $A$ is the cross-section area and  the $z$ coordinates of the top and bottom boundaries are given, as a retarded function of $t_0$, by 
\begin{equation}
\begin{array}{l}
{z_{Top}}\left( {{t_0}} \right) = \frac{1}{2}L + \varepsilon L\cos \left( {\omega {t_0} - k{r_0} + \frac{1}{2}{n_z}kL} \right)\\
{z_{Bottom}}\left( {{t_0}} \right) =  - \frac{1}{2}L - \varepsilon L\cos \left( {\omega {t_0} - k{r_0} - \frac{1}{2}{n_z}kL} \right)
\end{array}
\end{equation}
Here $k$ is the wave number $k=\omega/c$.

 The few grams of material in the boundary regions $z_{Bottom}(t')<z<-L/2$ and $L/2<z<Z_{Top}(t')$ contribute as much to the gravitational field as the hundreds of kilogram material in the interior. 
 
      The time-dependent component of $\bar h^{00}$ consists of integrals over the interior and over the boundaries
\begin{equation}
\begin{array}{c}
{{\bar h}^{00}}\left( {{\rm{interior}}} \right) =  - \frac{{4G}}{{{r_0}{c^2}}}\varepsilon {\rho _0}\pi A\int\limits_{ - L/2}^{L/2} {dz} \cos \left( {\frac{{\pi z}}{L}} \right)\cos \left( {\omega {t_0} - k{r_0} + {n_z}kz} \right) = \\
 =  - \frac{{8G}}{{{r_0}{c^2}}}M\varepsilon \cos \Phi \cos \left( {\frac{1}{2}{k_z}L} \right)\frac{{{\pi ^2}}}{{{\pi ^2} - {L^2}k_z^2}}
\end{array}
\label{interior}
\end{equation}
\begin{equation}
{\bar h^{00}}\left( {{\rm{boundaries}}} \right) = \frac{{4G}}{{{r_0}{c^2}}}{\rho _0}A\left[ {\int\limits_{{z_{Bottom}}\left( {{t_0}} \right)}^{ - L/2} {dz}  + \int\limits_{L/2}^{{z_{Top}}({t_0})} {dz} } \right] = \frac{{8G}}{{{r_0}{c^2}}}M\varepsilon \cos \Phi \cos \left( {\frac{1}{2}{k_z}L} \right)
\label{boundary}
\end{equation}
where $\Phi$ denotes for brevity the phase $\Phi=(\omega t_0 - k r_0)$ and $k_z$ is the $z$ component of the wave vector ${\bf k}=k{\bf n}$.

The boundary and interior terms have opposite signs, but comparable magnitudes.  Their sum is
\begin{equation}
{\bar h^{00}} = \frac{{8G}}{{{r_0}{c^2}}}M\varepsilon \cos \Phi \cos \left( {\frac{1}{2}{k_z}L} \right)\left( {1 - \frac{{{\pi ^2}}}{{{\pi ^2} - {L^2}k_z^2}}} \right) \cong - \frac{{8G}}{{{r_0}{c^2}}}M\varepsilon \cos \Phi \cos \left( {\frac{1}{2}{k_z}L} \right)\frac{{{L^2}k_z^2}}{{{\pi ^2}}}
\label{eq16}
\end{equation}

Note that the factor $\frac{{{L^2}k_z^2}}{{{\pi ^2}}}$
   is of second order in the ratio of the length of the Weber bar (of the order of 1 meter) and the wavelength of gravitational radiation (of the order of 100 km).  Also note that    Weinberg (\cite{Weinberg}, p.\ 270) appends alternating positive and negative masses to the ends of the Weber bar.  The mathematical justification for this procedure is unclear. The contributions to the metric from interior and boundary terms are comparable.  

Eq.\ (\ref{eq16}) can be rewritten
\begin{equation}
{\bar h^{00}}\left( {{{\bf{x}}_0},{t_0}} \right) =  - \frac{{8G}}{{{r_0}{c^4}}}M{L^2}{\omega ^2}n_z^2\frac{\varepsilon }{{{\pi ^2}}}\cos \left( {\omega {t_0} - k{r_0}} \right)
\end{equation}
In conclusion, the trace-reversed metric deviation from the retarded integral solution to the linearized Einstein equation in the harmonic gauge can be expressed in matrix form as follows
\begin{equation}
\left[ {\bar h^{\mu \nu , H}}\left( {{{\bf{x}}_0},{t_0}} \right) \right]=  - \frac{{8GM}}{{\pi^2{r_0}{c^4}}}{L^2}\varepsilon {\omega ^2}\left( {\begin{array}{*{20}{c}}
{n_z^2}&0&0&{{n_z}}\\
0&0&0&0\\
0&0&0&0\\
{{n_z}}&0&0&1
\end{array}} \right)\cos \left( {\omega {t_0} - k{r_0}} \right)
\label{h-H}
\end{equation}
The non-trace-reversed form is
\begin{equation}
\left[{h^{\mu \nu ,H}}\left( {{{\bf{x}}_0},{t_0}} \right) \right] =  - \frac{{4GM}}{{{\pi ^2}{r_0}{c^4}}}{L^2}\varepsilon {\omega ^2}\left( {\begin{array}{*{20}{c}}
{n_z^2 + 1}&0&0&{2{n_z}}\\
0&{n_z^2 - 1}&0&0\\
0&0&{n_z^2 - 1}&0\\
{2{n_z}}&0&0&{n_z^2 + 1}
\end{array}} \right)\cos \left( {\omega {t_0} - k{r_0}} \right)
\end{equation}

A metric of this kind is termed ``Class I'' by Boardman and Bergmann \cite{Boardman}. 

\section{Transverse traceless metric }
\label{tran}

In order to construct a spherical transverse-traceless metric we follow the recipe given by Finn \cite{Finn}. This is based on the quadrupole moment $Q_{ij}(t)$ of the mass distribution in the source ($i,j=1,2,3$)
\begin{equation}
{Q_{ij}}\left( t \right) = \int {{d^3}x\rho \left( {x,t} \right)\left( {{x_i}{x_j} - \frac{1}{3}{x^k}{x_k}{\delta _{ij}}} \right)} 
\end{equation}
The tensor ${Q_{ij}}\left( t \right)$ is combined with the projection operator ${P_{jk}}\left( {{\bf x_0}} \right) = {\delta _{jk}} - {n_j}{n_k}$, depending on the observation point ${\bf x}_0$, to obtain a tensor ${Q_{ij}^{TT}}\left( {\bf x}_0,t \right)$ 
\begin{equation}
Q_{ij}^{TT}\left( {\bf x}_0,t \right) = {P_{ik}}({\bf x_0}){Q_{kl}}{P_{lj}}\left( {{\bf x_0}} \right) - \frac{1}{2}{P_{ij}}\left( {{\bf x_0}} \right){Q_{lm}}{P_{lm}}\left( {{\bf x_0}} \right)
\end{equation}
The transverse traceless metric ${h_{ij}^{TT}}\left( {{\bf x_0},{t_0}} \right)$ is proportional to the second derivative of ${Q_{ij}^{TT}}\left( {\bf x}_0,t \right)$ with respect to $t$ evaluated at the retarded time $\left( t_0 - r_0/c \right)$
\begin{equation}
h_{ij}^{TT}\left( {{\bf x_0},{t_0}} \right) = \frac{{2G}}{{{r_0}{c^4}}}{\left[ {\frac{{{d^2}}}{{d{t^2}}}Q_{ij}^{TT}\left( {{{\bf{x}}_0},t} \right)} \right]_{t = {t_0} - {r_0}/c}}
\end{equation}

For our oscillating Weber bar the tensor $Q_{ij}(t) \equiv I_{ij}(t) - \frac{1}{3} I_{kk}(t)\delta_{ij}$ has one time-dependent component, namely
\begin{equation}
{I_{33}}\left( t \right) = \int {dA} \int\limits_{ - z\left( t \right)/2}^{z\left( t \right)/2} {dz} {z^2}\rho \left( {x,t} \right) = \frac{{M{L^2}}}{{12}} + M{L^2}\varepsilon \left( {\frac{4}{{{\pi ^2}}}} \right)\cos \omega t
\end{equation}
Note that evaluating this integral requires the same boundary/interior treatment as that used for calculating ${\bar h^{00}}$.

After multiplying matrices and considering that the components $h_{00}^{TT}$ and $h_{i0}^{TT}$ are zero due to the gauge condition, the result is
\begin{equation}
\begin{array}{l}
\left[ h_{\mu \nu }^{TT}({{\bf{x}}_0},{t_0}) \right]= \\
 - \frac{{4GM}}{{{\pi ^2}{r_0}{c^4}}}{L^2}\varepsilon {\omega ^2}\left( {\begin{array}{*{20}{c}}
0&0&0&0\\
0&{n_x^2n_z^2 - n_y^2}&{{n_x}{n_y}\left( {n_z^2 + 1} \right)}&{ - {n_x}{n_z}\left( {1 - n_z^2} \right)}\\
0&{{n_x}{n_y}\left( {n_z^2 + 1} \right)}&{n_y^2n_z^2 - n_x^2}&{ - {n_y}{n_z}\left( {1 - n_z^2} \right)}\\
0&{ - {n_x}{n_z}\left( {1 - n_z^2} \right)}&{ - {n_y}{n_z}\left( {1 - n_z^2} \right)}&{{{\left( {1 - n_z^2} \right)}^2}}
\end{array}} \right)\cos \left( {\omega {t_0} - k{r_0}} \right)
\end{array}
\label{h-TT}
\end{equation}

\subsection{Gauge transformation}
\label{gaug}

There exists no rigorous proof that the TT metric (\ref{h-TT}) can be related to the harmonic metric (\ref{h-H}) through a gauge transformation valid in all space. 
It is well known that the TT gauge can be imposed on a plane gravitational wave, leaving only the physical transverse components of the field. The same can be done for fields which are superpositions of plane waves, but the TT gauge is not obtainable in general, as we show in the following. 

The standard argument proving that it is possible to pass from the harmonic gauge to the TT gauge outside a source goes as follows \cite{Maggiore}. Suppose to have a trace-reversed metric  $\bar h^{\mu \nu} (x )$ which satisfies the harmonic gauge condition  ${\partial ^\mu }{\bar h_{\mu \nu }}(x) = 0$ and the field equation ${\partial ^\alpha }{\partial _\alpha }{\bar h_{\mu \nu }}(x) = 0$. The infinitesimal gauge transformations of the coordinates have the form
\begin{equation}
x{'^\mu } = {x^\mu } + {\xi ^\mu }(x)
\label{infinitesimal}
\end{equation}
 Under these transformations the metric changes as
\begin{equation}
\bar h{'_{\mu \nu }}(x') = {\bar h_{\mu \nu }}(x) - {\xi _{\mu \nu }}(x)
\label{trasf-h}
\end{equation}
where ${\xi _{\mu \nu }}(x)$ is defined as
\begin{equation}
{\xi _{\mu \nu }}(x) = {\partial _\mu }{\xi _\nu }(x) + {\partial _\nu }{\xi _\mu }(x) - {\eta _{\mu \nu }}{\partial _\rho }{\xi ^\rho }(x)
\label{def-xi2}
\end{equation}
The four-divergence of the metric changes as 
\begin{equation}
\left( {{\partial ^\mu }{{\bar h}_{\mu \nu }}(x)} \right)' = {\partial ^\mu }{\bar h_{\mu \nu }}(x) - {\partial ^\alpha }{\partial _\alpha }{\xi _\nu }(x)
\end{equation}
therefore the harmonic gauge is not spoiled in the transformation, provided
${\partial ^\alpha }{\partial _\alpha }{\xi _\nu }(x)=0$. By suitably choosing four functions ${\xi _\nu }(x)$  which satisfy this latter condition, and inserting them into (\ref{trasf-h}), (\ref{def-xi2}), we can hope to cancel the trace and some components of  $\bar h^{\mu \nu} (x )$, eventually obtaining the TT gauge. Relation (\ref{trasf-h}), however, is not just an algebraic equation, but a partial differential equation (except in some simple cases, like for plane waves). This introduces some complications.

Usually one starts by requiring that the trace of   $\bar h^{'\mu \nu} (x )$ is zero and then attempts to change the  $\bar h^{'0 i} $  components. The null-trace condition translates into the following condition on $\xi_\nu(x)$ 
\begin{equation}
{\eta ^{\mu \nu }}\bar h{'_{\mu \nu }}(x) = 0{\rm{  }} \Rightarrow {\rm{  }}{\eta ^{\mu \nu }}{\bar h_{\mu \nu }}(x) - {\eta ^{\mu \nu }}{\xi _{\mu \nu }}(x) = 0{\rm{  }} \Rightarrow {\rm{  }}2{\partial _\rho }{\xi ^\rho }(x) =  - {\eta ^{\mu \nu }}{\bar h_{\mu \nu }}(x)
\end{equation}
So, in order for $\bar h^{'\mu \nu} (x )$ to be traceless, the gauge transformation must satisfy the equations (in units such that $c=1$)
\begin{equation}
\left\{ \begin{array}{l}
{\partial _0}^2{\xi _\mu }(x) - \Delta {\xi _\mu }(x) = 0\\
{\partial _0}{\xi ^0}(x) - {\partial _i}{\xi ^i}(x) =  - \frac{1}{2}{\eta ^{\mu \nu }}{{\bar h}_{\mu \nu }}(x)
\end{array} \right.
\label{traceless}
\end{equation}
The solutions of these partial differential equations must respect some conditions at the boundary of the source, because inside the source it is impossible to pass to the TT gauge (${\partial ^\alpha }{\partial _\alpha }{\xi _\nu }(x)=0$  is not true there. See Fig.\ \ref{boundary}.)

\begin{figure}
\begin{center}
  \includegraphics[width=10cm,height=7cm]{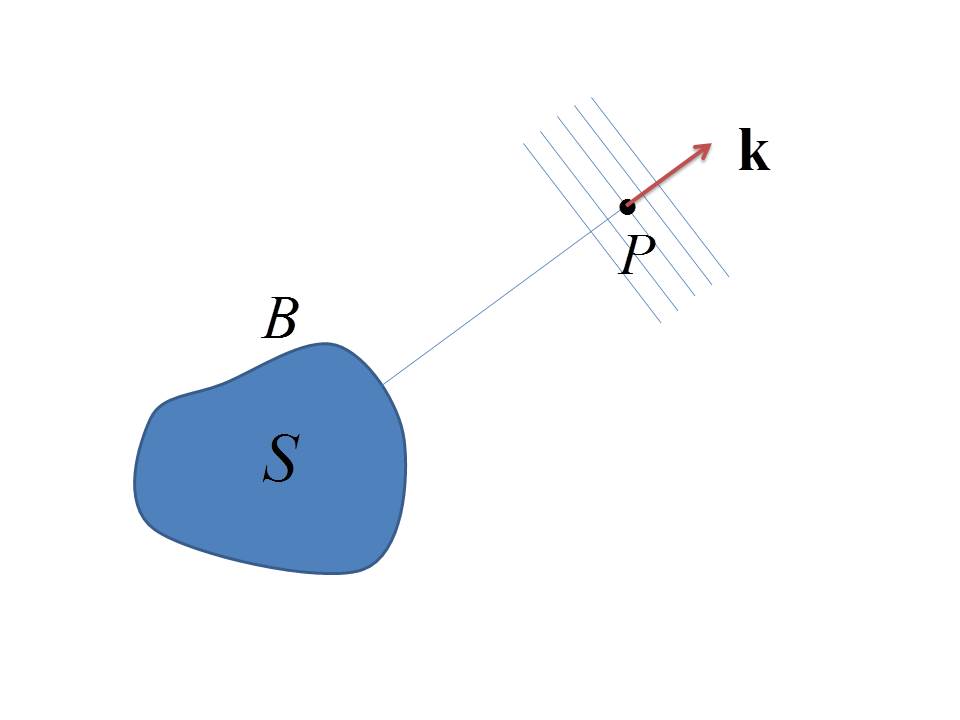}
\caption{Local implementation of the TT gauge. At any point $P$ in the radiation region the gravitational wave can be locally approximated with a plane wave of vector ${\bf k}$. The components $\chi_i$ of the gauge transformation to the TT gauge (see eq.\ (\ref{eq-chi})) are then chosen of the form $\chi_i({\bf x})=\theta_i e^{i {\bf k}{\bf x}}$, with $\theta_i k_i=0$. In order to obtain a global implementation of the gauge condition, however, we should solve eq.\ (\ref{eq-chi}) over all space outside the source $S$, with suitable boundary conditions at the border $B$ of $S$.
} 
\label{boundary}
\end{center}      
\end{figure}

Even disregarding the boundary conditions, it can be impossible to find a solution of (\ref{traceless}) in a general case. In the wave region the trace of the (given) harmonic metric has time dependence of the form 
${\eta ^{\mu \nu }}{\bar h_{\mu \nu }}(x) = f({\bf{x}}){e^{i\omega t}}$, so we make the ansatz ${\xi _\rho }({\bf{x}},t) = {\chi _\rho }({\bf{x}}){e^{i\omega t}}$   and the second equation of (\ref{traceless}) becomes an equation for ${\chi _\rho }({\bf{x}})$ 
\begin{equation}
i\omega {\chi _0}({\bf{x}}) - {\partial _i}{\chi ^i}({\bf{x}}) =  - \frac{1}{2}f({\bf{x}})
\label{eq-chi}
\end{equation}
We can solve this by requiring that ${\partial _i}{\chi ^i}({\bf{x}}) = 0$  and ${\chi _0}({\bf{x}}) =  - \frac{1}{{2i\omega }}f({\bf{x}})$. 
There is a trivial solution ${\chi ^i}({\bf{x}}) = 0$ , but this certainly does not allow to cancel the components ${\bar h_{0i}}$  in the next step. If we take ${\chi ^i}({\bf{x}})$  to be a transverse plane wave with wave vector ${\bf{k}}$  we can satisfy the condition ${\partial _i}{\chi ^i}({\bf{x}}) = 0$   and then also the second equation of (\ref{traceless}), provided $|{\bf{k}}|=\omega$;  but after that, we will be able to cancel the components ${\bar h_{0i}}$  only if they are plane waves themselves. In spherical symmetry we can satisfy the condition   ${\partial _i}{\chi ^i}({\bf{x}}) = 0$ with functions ${\chi ^i}({\bf{x}}) \neq 0$  which are like the components of an electric field in vacuum and behave like $1/ {\bf{x}}^2$; but at the same time these functions should satisfy the second condition of (\ref{traceless}), namely $ - {\omega ^2}{\chi _i}({\bf{x}}) - {\partial ^k}{\partial _k}{\chi _i}({\bf{x}}) = 0$, and this is clearly impossible.

Nevertheless, the gauge transformation holds locally because it is always possible to approximate the wave with a plane wave in a region much smaller than its wavelength. This provides an indirect proof that the Riemann tensors for the harmonic gauge and the TT gauge are the same at any point (an explicit calculation is given in Sect.\ \ref{off}). This ensures, in turn, that the physical effects of the field on detectors are the same, as expected. Nevertheless, in the following sections we analyze in more detail the behavior of gravitational waves detectors also in more general situations, where the TT gauge is not applicable.

\section{Gravitational forces on a detector: basic definitions }
\label{det}

Gravitational forces on a detector come in two varieties.  For observation points in the near field, the metric varies over distances comparable to the distance from source to observer.  In the far field, the metric varies over distances comparable to the wavelength of radiation.  
     For measurements in the near field, gravitational forces can be expressed in terms of the geodesic equation
\begin{equation}
\frac{{{d^2}{x^\mu }}}{{d{\tau ^2}}} =  - \Gamma _{{\rm{ }}\nu \rho }^\mu \frac{{d{x^\nu }}}{{d\tau }}\frac{{d{x^\rho }}}{{d\tau }}
\end{equation}
where  $x^\mu(\tau)$ describes the trajectory of a test particle, and the Christoffel symbol describing variations in the metric are defined as
\begin{equation}
\Gamma _{{\rm{  }}\sigma \nu }^\pi  = {g^{\pi \rho }}\left[ {\sigma \nu ,\rho } \right];{\rm{ \ \ \ \ \       }}\left[ {\sigma \nu ,\rho } \right] = \frac{1}{2}\left( {{g_{\nu \rho ,\sigma }} + {g_{\rho \sigma ,\nu }} - {g_{\sigma \nu ,\rho }}} \right)
\end{equation}
Forces on a stationary test particle can be described in terms of a gravitoelectric field
\begin{equation}
E_g^i =  - \Gamma _{00}^i{c^2} = \frac{{c\partial {h^{0i}}}}{{\partial t}} + \frac{{{c^2}\partial {h^{00}}}}{{2\partial {x^i}}}
\label{ge}
\end{equation}
At larger distances from the source, tidal forces associated with time derivatives of the spatial components $h^{ij}$  are important.  The geodesic distance between two points ${\bf x}_0$  and ${\bf x}$  in a detector differs from the coordinate distance.  Since electromagnetic forces and wavefunctions are constrained to propagate along the geodesic distance, the two points must have a relative acceleration to balance stresses and strains
 \begin{equation}
\frac{{{d^2}{X_i}}}{{d{\tau ^2}}} = \frac{1}{2}\frac{{{\partial ^2}{h_{ij}}}}{{\partial {\tau ^2}}}{X^j}
\label{tidal-forces}
\end{equation}
where ${\bf X}={\bf x}_0-{\bf x}$  is the coordinate separation of the two points.  The transition from near field to far field behavior is expressed smoothly in terms of the geodesic deviation equation, which involves the Riemann tensor,  
 \begin{equation}
\frac{{{D^2}\delta {x^\lambda }}}{{D{\tau ^2}}} =  - R_{{\rm{  }}\nu \mu \rho }^\lambda \delta {x^\mu }\frac{{d{x^\nu }}}{{d\tau }}\frac{{d{x^\rho }}}{{d\tau }}
 \end{equation}
where $D$ denotes a covariant derivative, $\delta {x^\lambda }$  denotes the coordinate difference between a spacetime point $x^\lambda$  and a reference point.  A convenient formula for the Riemann tensor in covariant form is given by 
\begin{equation}
{R_{\rho \sigma \mu \nu }} = {\left[ {\sigma \nu ,\rho } \right]_{,\mu }} - {\left[ {\sigma \mu ,\rho } \right]_{,\nu }} + \Gamma _{{\rm{  }}\sigma \mu }^\pi \left[ {\rho \nu ,\pi } \right] - \Gamma _{{\rm{  }}\sigma \nu }^\pi \left[ {\rho \mu ,\pi } \right]
\end{equation}
     Ignoring the higher order terms in the Riemann tensor, the relative acceleration between neighboring points can be approximated as follows
\begin{equation}
\frac{{{D^2}\delta {x^i}}}{{D{\tau ^2}}} \cong  - R_{{\rm{  0j}}0}^i{c^2}\delta {x^j}
\end{equation}
The four-velocity of a point is approximated in terms of the time-like component
\begin{equation}
\frac{{d{x^\mu }}}{{d\tau }} = \gamma \left( {c,{v_x},{v_y},{v_z}} \right) \cong \left( {c,0,0,0} \right)
\end{equation}
The velocity of the detector is assumed to be small compared to the speed of light.  

In this work the relation between gravitoelectric and tidal forces is displayed explicitly, for the radiation field of the Weber bar.  In the linearized approximation, using the expression for the TT metric, the gravitoelectric field defined in (\ref{ge}) is given by the vector 
\begin{equation}
E_g^i =  - \frac{{4GM}}{{{\pi ^2}r{c^3}}}{L^2}\varepsilon {\omega ^3}\sin \left( {\omega t - kr} \right)\left( {{n_x}\left( {n_z^2 + 1} \right),{n_y}\left( {n_z^2 + 1} \right),\frac{1}{2}n_z^3 - \frac{3}{2}{n_z}} \right)
\label{e-i-g}
\end{equation}
It is not obvious at this stage that the Riemann tensor nullifies the gravitoelectric field in radiation.

\subsection{Gravitational wave detector on the x axis}
\label{x-axis}
 
\begin{figure}
\begin{center}
  \includegraphics[width=10cm,height=7cm]{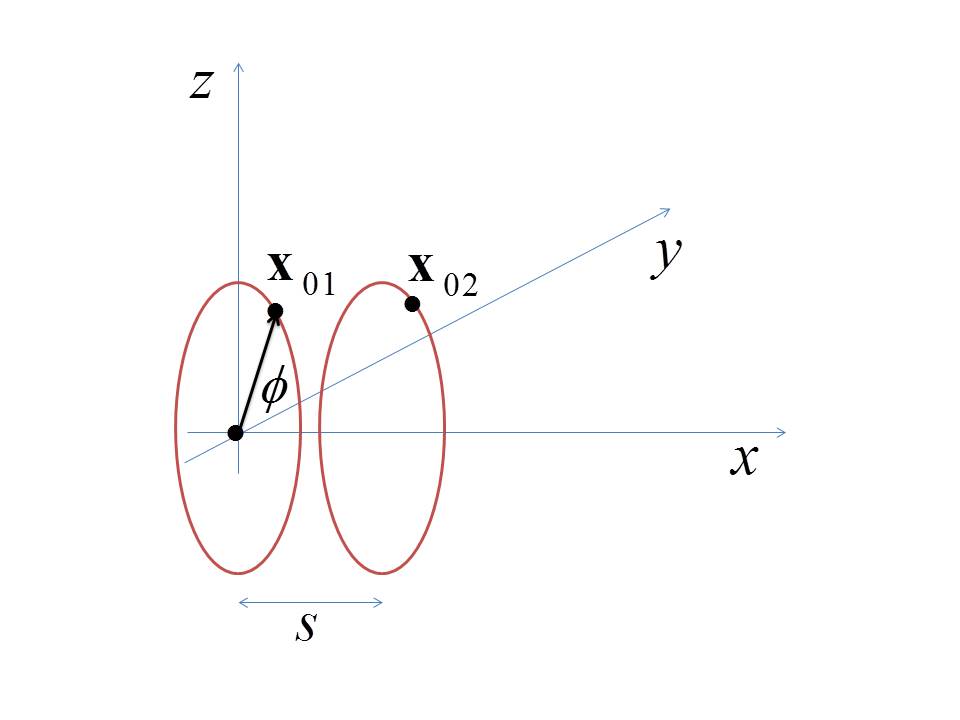}
\caption{General detector for gravitational waves. The detector consists of two rings centered on the $x$ axis.  The rings each have a radius $B$, oriented parallel to the $yz$ plane.  In the absence of gravitational radiation, the distance between the rings is $s$.} 
\label{rings1}
\end{center}
      
\end{figure}

     Having explicit expressions for the harmonic metric (\ref{h-H}) and the TT metric (\ref{h-TT}) allows gravitoelectric and strain field forces on a detector to be evaluated. The detector consists of two rings centered on the $x$ axis.  The rings each have a radius $B$, oriented parallel to the $yz$ plane.  In the absence of gravitational radiation, the distance between the rings is $s$.  
   Anticipating the need to evaluate phase gradients, define the origin of coordinates as the center of Ring 1 (see Fig.\ \ref{rings1}).  For an observer on the $x$ axis, the TT strain field is given by 
\begin{equation}
h_{ij}^{TT}\left( {{n_x} = 1} \right) =  - H\cos \left( {\omega t - k{r_1} - kx} \right)\left( {\begin{array}{*{20}{c}}
0&0&0\\
0&{ - 1}&0\\
0&0&1
\end{array}} \right)
\end{equation}
where
\begin{equation}
H = \frac{{8GM}}{{{\pi ^2}r{c^4}}}{L^2}\varepsilon {\omega ^2}
\end{equation}

Consider points in rings 1 and 2 originally at the locations 
\begin{equation}
\begin{array}{l}
{{\bf x}_{01}} = \left( {0,B\cos \phi ,B\sin \phi } \right)\\
{{\bf x}_{02}} = \left( {s,B\cos \phi ,B\sin \phi } \right)
\end{array}
\end{equation}

The tidal forces (eq.\ (\ref{tidal-forces})) produce the following accelerations
\begin{equation}
\begin{array}{c}
\ddot x_1^{TT} = \frac{1}{2}H{\omega ^2}\left( {\begin{array}{*{20}{c}}
0&0&0\\
0&{ - 1}&0\\
0&0&1
\end{array}} \right)\left( {\begin{array}{*{20}{c}}
0\\
{B\cos \phi }\\
{B\sin \phi }
\end{array}} \right)\cos \left( {\omega t - k{r_1}} \right)  \\
=\frac{1}{2}HB{\omega ^2}\left( {0, - \cos \phi ,\sin \phi } \right)\cos \left( {\omega t - k{r_1}} \right)
\end{array}
\end{equation}
\begin{equation}
\ddot x_2^{TT} = \frac{1}{2}HB{\omega ^2}\left( {0, - \cos \phi ,\sin \phi } \right)\cos \left( {\omega t - k{r_1} - ks} \right)
\end{equation}
In particular, the longitudinal force is zero.  
For the harmonic metric, the strain field is 
\begin{equation}
h_{ij}^H\left( {{n_x} = 1} \right) =  - H\cos \left( {\omega t - k{r_1} - kx} \right)\left( {\begin{array}{*{20}{c}}
{ - 1}&0&0\\
0&{ - 1}&0\\
0&0&1
\end{array}} \right)
\end{equation}
The tidal forces produce the following accelerations
\begin{equation}
\begin{array}{c}
\ddot x_1^H = \frac{1}{2}H{\omega ^2}\left( {\begin{array}{*{20}{c}}
{ - 1}&0&0\\
0&{ - 1}&0\\
0&0&1
\end{array}} \right)\left( {\begin{array}{*{20}{c}}
0\\
{B\cos \phi }\\
{B\sin \phi }
\end{array}} \right)\cos \left( {\omega t - k{r_1}} \right)\\
 = \frac{1}{2}HB{\omega ^2}\left( {0, - \cos \phi ,\sin \phi } \right)\cos \left( {\omega t - k{r_1}} \right)
\end{array}
\end{equation}
\begin{equation}
\begin{array}{c}
\ddot x_2^H = \frac{1}{2}H{\omega ^2}\left( {\begin{array}{*{20}{c}}
{ - 1}&0&0\\
0&{ - 1}&0\\
0&0&1
\end{array}} \right)\left( {\begin{array}{*{20}{c}}
s\\
{B\cos \phi }\\
{B\sin \phi }
\end{array}} \right)\cos \left( {\omega t - k{r_1} - ks} \right)\\
 = \frac{1}{2}H{\omega ^2}\left( { - s, - B\cos \phi ,B\sin \phi } \right)\cos \left( {\omega t - k{r_1} - ks} \right)
\end{array}
\label{tidal-force}
\end{equation}
The tidal force includes a longitudinal component, absent from the TT force.  
    In addition, the harmonic metric includes a gravitoelectric field from eq.\ (\ref{e-i-g})  
\begin{equation}
{\bf E}_g =  - \frac{1}{2}H\omega c\left( {1,0,0} \right)\sin \left( {\omega t - k{r_1} - kx} \right)
\end{equation}
The phase difference causes a relative acceleration between the rings. We denote this acceleration with the superscript ``$g$'', in order to distinguish it from the ``strain'' accelerations computed above. The superscript ``$H$'' means, as above, that this acceleration is computed in the harmonic gauge.
\begin{equation}
\begin{array}{c}
\ddot x_2^{H,g} - \ddot x_1^{H,g} =  - \frac{1}{2}H\omega c\left( {\sin \left( {\omega t - k{r_1} - ks} \right) - \sin \left( {\omega t - k{r_1}} \right)} \right)\\
 = \frac{1}{2}H{\omega ^2}s\cos \left( {\omega t - k{r_1} - ks} \right) - \frac{1}{2}H\omega c\left[ {\frac{3}{2}{{\left( {ks} \right)}^2}\sin \left( {\omega t - k{r_1}} \right) + O\left( {{{\left( {ks} \right)}^3}} \right)} \right]
\end{array}
\label{ge-accel}
\end{equation}
The relative acceleration produced by the gravitoelectric field (\ref{ge-accel}) cancels the longitudinal component of the tidal force (eqn.\ (\ref{tidal-force})), up to second order in the phase $ks$.   

Both the harmonic and TT metrics produce transverse motion of the detector rings and surrounding instrumentation.  The harmonic metric also produces a longitudinal oscillation of detector and instrumentation.  However the sum of the longitudinal forces from the gravitoelectric field gradient and strain fields cancel in first order in the phase $ks$, producing a small net oscillation in the separation of particles.  The expression of $T^{33}$ in eq.s (\ref{T-Lagr}), (\ref{T-Eul}) for stress in a material requires a strain gradient.   Since only relative motion is detectable, the harmonic metric produces almost the same detector response as the TT metric.    

     The geodesic deviation equation relates strains in the $x$ direction to Riemann tensor elements of the form $R_{10j0}$.  
\begin{equation}
{R_{1010}} = {\left[ {00,1} \right]_{,1}} - {\left[ {01,1} \right]_{,0}} =  - \frac{1}{2}{g_{00,11}} - \frac{1}{2}{g_{11,00}}
\end{equation}
\begin{equation}
\left[ {00,1} \right] = \frac{1}{2}\left( {{g_{01,0}} + {g_{10,0}} - {g_{00,1}}} \right) =  - \frac{1}{2}{g_{00,1}}
\end{equation}
\begin{equation}
\left[ {01,1} \right] = \frac{1}{2}\left( {{g_{11,0}} + {g_{10,1}} - {g_{01,1}}} \right) = \frac{1}{2}{g_{11,0}}
\end{equation}
\begin{equation}
{R_{1020}} = 0; \ \ \ \ {R_{1030}} = 0
\end{equation}

For the harmonic metric
\begin{equation}
{g_{00}} =  - {g_{11}} =  - H\cos \left( {\omega t - k{r_1} - kx} \right)
\end{equation}
Hence
\begin{equation}
{g_{00,11}} + {g_{11,00}} = \frac{{{\partial ^2}{g_{00}}}}{{\partial {x^2}}} + \frac{{{\partial ^2}{g_{11}}}}{{{c^2}\partial {t^2}}} = H\cos \left( {\omega t - k{r_1}{\kern 1pt}  - kx} \right)\left( { - {k^2} + \frac{{{\omega ^2}}}{{{c^2}}}} \right) = 0
\end{equation}
since $\omega=kc$.  

The derivative $g_{00,1}$  is the $x$ component of the gravitoelectric field.  The gradient of the gravitoelectric field cancels the second time derivative of $h_{11}$.  Since the $R_{10j0}$ terms of the Riemann tensor vanish, the $x$ component of the strain vanishes.

\subsection{Gravitational wave detector on the z axis}
\label{z-axis}

Consider the detector in Fig.\ \ref{rings1} located on the $z$ axis.  At this location, the TT metric (\ref{h-TT}) vanishes, and the harmonic gravitoelectric field (\ref{e-i-g}) is non-zero.  A longitudinal component of the harmonic tidal strain field persists, resulting in oscillations between the two rings.  
\begin{equation}
\ddot z_2^{H,strain} - \ddot z_1^{H,strain} = \frac{1}{2}H{\omega ^2}\left( {\begin{array}{*{20}{c}}
0&0&0\\
0&0&0\\
0&0&1
\end{array}} \right)\left( {\begin{array}{*{20}{c}}
{B\cos \phi }\\
{B\sin \phi }\\
s
\end{array}} \right)\cos \left( {\omega t - k{r_1} - ks} \right)
\end{equation}
The effect of the gravitoelectric field on the difference $z_2-z_1$ is as follows
\begin{equation}
\begin{array}{c}
\ddot z_2^{H,g} - \ddot z_1^{H,g} = \frac{1}{2}H\omega c\left( {\sin \left( {\omega t - k{r_1} - ks} \right) - \sin \left( {\omega t - k{r_1}} \right)} \right) = \\
 - \frac{1}{2}H{\omega ^2}s\cos \left( {\omega t - k{r_1} - ks} \right) + \frac{1}{2}H\omega c\left[ {\frac{1}{2}{{\left( {ks} \right)}^2}\sin \left( {\omega t - k{r_1}} \right) + O\left( {{{\left( {ks} \right)}^3}} \right)} \right]
\end{array}
\end{equation}
The gravitoelectric field produces longitudinal motion of the detector and surrounding instrumentation, which cannot be detected with local measurements.  The gradient of the gravitoelectric field cancels the tidal strain, up to second order in the phase $ks$.  The relevant Riemann tensor element is 
\begin{equation}
{R_{3030}} = {g_{30,03}} - \frac{1}{2}{g_{00,33}} - \frac{1}{2}{g_{33,00}}=0
\end{equation}
since
\begin{equation}
{g_{00}} =  - {g_{30}} = {g_{33}} =  - H\cos \left( {\omega t - k{r_1} - kz} \right)
\end{equation}
The combination $ - {g_{30,0}} + \frac{1}{2}{g_{00,3}}$  is the gravitoelectric field.  The gradient of the gravitoelectric field cancels the second time derivative of $g_{33}$.  Since the $R_{3030}$ Riemann tensor element vanishes, the strain force in the $z$ direction is zero.

\section{Time-dependent near fields  }
\label{near}

One technique for computing near fields from retarded integrals (\ref{general-ret-integral}) is to use series expansions.  The denominator can be expanded as in eq.\ (\ref{series1}). The time retardation involves the distance between source and observer, eq.\ (\ref{series2}). 
The result is a double infinite series of terms, which can be evaluated using trigonometric identities and integration by parts.
     Retaining only time-dependent terms up to second order in wavenumber $k$ and third order in $1/r_0$, the trace-reversed metric deviation tensor is as follows 
\begin{equation}
\bar h^{H,\mu \nu }\left( {{{\bf x}_0},{t_0}} \right) = \frac{{8GM{L^2}\varepsilon }}{{{\pi ^2}{c^2}}}\left( {\begin{array}{*{20}{c}}
\begin{array}{l}
\left( {\frac{{ - k_z^2}}{{{r_0}}} + \frac{{\left( {1 - 3n_z^2} \right)}}{{r_0^3}}} \right)\cos \Phi \\
 + \frac{{k\left( {1 - n_z^2} \right)}}{{r_0^2}}\sin \Phi 
\end{array}&0&0&{\frac{{ - k{k_z}}}{{{r_0}}}\cos \Phi  + \frac{{{k_z}}}{{r_0^2}}\sin \Phi }\\
0&0&0&0\\
0&0&0&0\\
{\frac{{ - k{k_z}}}{{{r_0}}}\cos \Phi  + \frac{{{k_z}}}{{r_0^2}}\sin \Phi }&0&0&{\frac{{ - {k^2}}}{{{r_0}}}\cos \Phi }
\end{array}} \right)
\label{eq60}
\end{equation}
where $\Phi=\left( \omega t_0 - k{r_0}  \right)$. As an example, the 03 metric element is related to the mass current
\begin{equation}
\bar h_{near}^{03}\left( {{{\bf{x}}_0},{t_0}} \right) = \frac{{4G}}{{r_0^2{c^3}}}\int {{d^3}x} J\left( {{\bf{x}},{t_0} - \frac{{{r_0}}}{c}} \right){\bf{n}} \cdot {\bf{x}}
\end{equation}
The equation involves the following integral
\begin{equation}
I_{near}^{03} = \int\limits_{ - L/2}^{L/2} {dz} \sin \left( {\frac{{\pi z}}{L}} \right){n_z}z = \frac{{{L^2}}}{{{\pi ^2}}}{n_z}\left[ {\sin u - u\cos u} \right]_{ - \pi /2}^{\pi /2} = \frac{{2{L^2}}}{{{\pi ^2}}}{n_z}
\end{equation}

\subsection{Gravitoelectric and gravitomagnetic fields }

     Forces on a test mass can be expressed in terms of the geodesic equation
\begin{equation}
\frac{{{d^2}{x^i}}}{{d{t^2}}} =  - \Gamma _{00}^i{c^2} - \Gamma _{0j}^ic{v^j} - \Gamma _{k0}^i{v^k}c
\end{equation}
Remembering that
\begin{equation}
\Gamma _{0j}^i = \frac{1}{2}{g^{ii}}\left( {\frac{{\partial {g_{ji}}}}{{\partial {x^0}}} + \frac{{\partial {g_{0i}}}}{{\partial {x^j}}} - \frac{{\partial {g_{0j}}}}{{\partial {x^i}}}} \right); \ \ \ \ \ 
\Gamma _{00}^i = \frac{1}{2}{g^{ii}}\left( {\frac{{\partial {g_{0i}}}}{{\partial {x^0}}} + \frac{{\partial {g_{0i}}}}{{\partial {x^0}}} - \frac{{\partial {g_{00}}}}{{\partial {x^i}}}} \right)
\label{geodesic}
\end{equation}
normalization and sign conventions are chosen so that the force on a test mass can be expressed as follows
\begin{equation}
{\bf{F}} = m{{\bf{E}}_g} + m{\bf{v}} \times {{\bf{B}}_g} - \frac{1}{2}m\frac{{\partial {h_{ij}}}}{{\partial t}}{v^i}{\hat \varepsilon _j}
\label{force}
\end{equation}
where ${\hat \varepsilon _j}$  is a unit vector in coordinate direction $j$.  
     The gravitoelectric field is expressed in terms of metric elements as follows
\begin{equation}
E_g^i =  - {c^2}\Gamma _{00}^i \cong c\frac{{\partial {h^{0i}}}}{{\partial t}} + \frac{1}{2}{c^2}\frac{{\partial {h^{00}}}}{{\partial {x^i}}} = \frac{1}{4}{c^2}\frac{{\partial {{\bar h}^{00}}}}{{\partial {x^i}}} + \frac{1}{4}{c^2}\frac{{\partial {{\bar h}^{33}}}}{{\partial {x^i}}} + c\frac{{\partial {h^{03}}}}{{\partial t}}\delta _3^i
\label{eq66}
\end{equation}
It is convenient to express the gravitomagnetic field as the curl of a vector potential
\begin{equation}
{{\bf{B}}_g} = \nabla  \times {\bf{h}} = c\left( { - {\partial _y}{h^{03}},{\partial _x}{h^{03}},0} \right)
\label{eq67}
\end{equation}
To conform to the geodesic equation, the vector potential must be chosen in terms of covariant metric elements
\begin{equation}
{\bf{h}} = c\left( {{h_{01}},{h_{02}},{h_{03}}} \right) =  - c\left( {{h^{01}},{h^{02}},{h^{03}}} \right)
\end{equation}
    For bookkeeping purposes, derivatives are labeled according to orders in $k$ and $1/r$.  The most important terms in the near field region involve gradients of the $1/r_0^3$  term of the $\bar h^{00}$  metric element
\begin{equation}
{{\bf{E}}_g}\left( {\frac{1}{{{r^4}}}} \right) \cong \frac{{2GM{L^2}\varepsilon }}{{{\pi ^2}{c^2}}}\cos \Phi \nabla \left[ {\frac{{\left( {1 - 3n_z^2} \right)}}{{r_0^3}}} \right]
\end{equation}
\begin{equation}
{{\bf{E}}_g}\left( {\frac{1}{{{r^4}}}} \right) \cong \frac{{6GM{L^2}\varepsilon }}{{{\pi ^2}{c^2}{r^7}}}\cos \Phi \left( {4x{z^2} - {y^2}x - {x^3},4y{z^2} - {x^2}y - {y^3},2{z^3} - 3{x^2}z - 3{y^2}z} \right)
\label{Eg-69}
\end{equation}
The leading term for the gravitomagnetic field involves the curl of the spatial factor in the $1/r_0^2$  term of the $\bar h^{03}$  metric element
\begin{equation}
{{\bf{B}}_g}\left( {\frac{k}{{{r^3}}}} \right) =  - \frac{{8GM{L^2}\varepsilon }}{{{\pi ^2}{c^2}}}k\sin \Phi \left[ {\nabla  \times \frac{{z\hat z}}{{{r^3}}}} \right] = \frac{{24GM{L^2}\varepsilon }}{{{\pi ^2}{c^2}}}\frac{{k\sin \Phi }}{{{r^5}}}\left( { - yz,xz,0} \right)
\label{eq70}
\end{equation}

\subsection{Displacement current }

    The displacement current involves a relation between the curl of the gravitomagnetic field and time derivative of the gravitoelectric field.  The leading term (linear in $k$) in the curl is 
\begin{equation}
\nabla  \times {{\bf{B}}_g}\left( {\frac{k}{{{r^4}}}} \right) = \frac{{24GM{L^2}k\varepsilon }}{{{\pi ^2}{c^2}{r^7}}}\left( {{x^3} + {y^2}x - 4x{z^2},{y^3} + {x^2}y - 4y{z^2},3{x^2}z + 3{y^2}z - 2{z^3}} \right)\sin \Phi 
\end{equation}
For comparison, the time derivative of the gravitoelectric field (\ref{Eg-69}) is 
\begin{equation}
\frac{{\partial {{\bf{E}}_g}}}{{ct}}\left( {\frac{k}{{{r^4}}}} \right) \cong  - \frac{{6GM{L^2}\varepsilon k}}{{{\pi ^2}{c^2}{r^7}}}\sin \Phi \left( {4x{z^2} - {y^2}x - {x^3},4y{z^2} - {x^2}y - {y^3},2{z^3} - 3{x^2}z - 3{y^2}z} \right)
\end{equation}
The two expressions are related:
\begin{equation}
\nabla  \times {{\bf{B}}_g}\left( {\frac{k}{{{r^4}}}} \right) = 4\frac{{\partial {{\bf{E}}_g}}}{{c\partial t}}\left( {\frac{k}{{{r^4}}}} \right)
\end{equation}
The relative signs in the displacement current equation are the same as the corresponding equation in electromagnetism.  However the relations between fields and sources (eqns. (\ref{eq66}) and (\ref{eq67})) have opposite signs to the corresponding relations in electromagnetism.
  
\subsection{Faraday induction}

Faraday induction involves a relation between the curl of the gravitoelectric and time derivative of the gravitomagnetic field. Since the curl of the gradients in eq.\ (\ref{eq66}) vanish, only the time derivative of the vector potential contributes to the curl of the gravitoelectric field.
\begin{equation}
{{\bf{E}}_g} = \nabla \left( {term} \right) + c\frac{{\partial {h^{03}}}}{{\partial t}}
\end{equation}
The equation for the gravitomagnetic field is 
\begin{equation}
{{\bf{B}}_g} = \nabla  \times {\bf{h}}
\end{equation}
     Since space and time derivatives commute, 
\begin{equation}
\nabla  \times {{\bf{E}}_g} =  - \frac{{\partial {{\bf{B}}_g}}}{{\partial t}}
\label{eq76}
\end{equation}
Note that ${h_z} = c{h_{03}} =  - c{h^{03}}$, resulting in the minus sign in eq.\ (\ref{eq76}).  Conclusion:  both Faraday induction and the displacement current obey laws similar to those for electromagnetism, in the near field region.  In vacuo the equations have the same sign structure as in electromagnetism.  However the relations  between fields and sources have opposite signs for gravitoelectromagnetism.  

What about the $h^{ij}$ derivative forces? 
     The force equation (\ref{force}) includes velocity-dependent forces on a test mass associated with time derivatives of spatial components of the metric.  The three diagonal elements of the metric deviation are as follows
\begin{equation}
{h^{11}} = {h^{22}} = \frac{1}{2}\left( {{{\bar h}^{00}} - {{\bar h}^{33}}} \right); \ \ \ \ \ {h^{33}} = \frac{1}{2}\left( {{{\bar h}^{00}} + {{\bar h}^{33}}} \right)
\end{equation}
All three elements include the term 
\begin{equation}
{h^{ii}} = \frac{{8GM{L^2}\varepsilon }}{{2{\pi ^2}{c^2}}}\frac{{\left( {1 - 3n_z^2} \right)}}{{{r^3}}}\cos \Phi  + ...
\end{equation}
The time derivative of these elements is of order  $k/r^3$   
\begin{equation}
\frac{{\partial {h^{ii}}}}{{c\partial t}}\left( {\frac{k}{{{r^3}}}} \right) =  - \frac{{8GM{L^2}\varepsilon }}{{2{\pi ^2}{c^2}}}\frac{{\left( {1 - 3n_z^2} \right)}}{{{r^3}}}k\sin \Phi 
\end{equation}
These terms are of the same order (linear in $k$) as that of the leading term (\ref{eq70}) in the gravitomagnetic field.

Concerning the relation between the harmonic gauge and the TT subgauge, remember that the linearized Einstein equation (\ref{eq1}) is derived by applying the harmonic gauge condition ${\partial _\mu }{\bar h^{\mu \nu }} = 0$,   to disentangle relations among metric elements.  As a consequence, derivatives of metric elements (\ref{eq60}) should be related; for example 
\begin{equation}
\frac{{\partial {{\bar h}^{00}}}}{{c\partial t}} + \frac{{\partial {h^{03}}}}{{\partial z}} = 0
\end{equation}
The $z$ derivative of the spatial factor in the $1/r_0$ term of $h^{03}$ results in a $1/r_0^2$ term.  This term is matched by the time derivative of the phase factor in the $1/r_0^2$  term of $\bar h^{00}$.  The Lorenz gauge condition requires relations among terms with differing powers of $1/r_0$. The radiation metrics (\ref{h-H}) and (\ref{h-TT}) lack the higher order terms to fulfill the harmonic gauge condition.  These metrics can be considered as solutions to the Einstein equation only in a restricted sense.  The metrics must be treated as if they were locally plane waves, so that derivatives of the $1/r_0$  factor are to be ignored.  The mathematics of this ansatz is described more carefully in section \ref{gaug}.

\subsection{Numerical integration}
\label{numerical}
 
      Another technique for evaluating retarded integrals is numerical integration.  As an example, the behavior of the gravitational field along the $z$ axis was computed using numerical integration.  Specifically, the integrals calculated numerically are as follows
\begin{equation}
\bar h_{num}^{\mu \nu }\left( {{z_0},{t_0}} \right) = \int\limits_{{z_{Bottom}}}^{{z_{Top}}} {dz} \frac{{{T^{\mu \nu }}\left( {z,t'} \right)}}{{\left| {{z_0} - z} \right|}} = \sum\limits_{i = 1}^{200} {\Delta z} \frac{{{T^{\mu \nu }}\left( {{z_i},t'} \right)}}{{{z_0} - {z_i}}}
\end{equation}
No series expansions were used, so that the main source of error is numerical precision, rather than truncation of a series.  

     The retarded time $t'$ assigned to each $z$ value in a coordinate grid is $t’=t_0-(z_0-z)/c$, with $t_0$ chosen as one tenth of an oscillation period.  The Weber bar was divided into 200 grid cells.  The perturbation parameter was chosen as $\varepsilon=0.005$.  For sufficient accuracy in computing $\bar h_{num}^{00}$ (interior), a Simpson rule integration over the width of a cell was used.  In Visual Basic code, 
\texttt{
H00=H00+H00cof*(RHmn+4*RHcn+RHpl)/6},
where \texttt{RHcn} is the mass density perturbation (\ref{eq14}) at the center of a cell, and \texttt{RHmn} and \texttt{RHpl} are density perturbations at the bottom and top of a cell. 

\begin{figure}
\begin{center}
  \includegraphics[width=10cm,height=7cm]{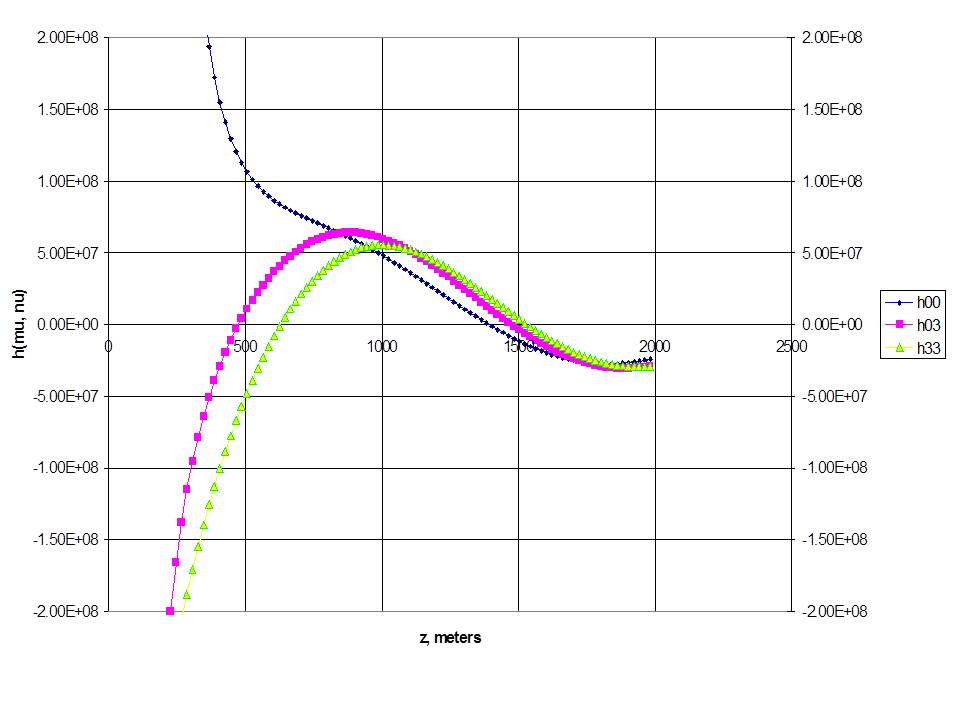}
\caption{Metric components in the transition region between near and far fields. At a distance of one wavelength, ${\bar h^{00}}$ merges smoothly with ${\bar h^{03}}$ and ${\bar h^{33}}$.  
     Close to an oscillating source, the time-varying gravitational disturbance is measurable.  For example, a torsion pendulum is used  to search for deviations from Newtonian gravity \cite{Schlamm}.  Beyond a distance of a wavelength, ${\bar h^{00}}$, ${\bar h^{03}}$ and ${\bar h^{33}}$ oscillate in unison, resulting in zero for the $R_{3030}$ Riemann tensor element. } 
\label{near-field}
\end{center}
      
\end{figure}

Showing the transition from near-field to far-field behavior for the real Weber bar is problematic, because the cancellation between ${\bar h^{00}}\left( {{\rm{interior}}} \right)$ (\ref{interior}) and ${\bar h^{00}}\left( {{\rm{boundary}}} \right)$ (\ref{boundary}) requires high numerical precision.  To obtain a numerically significant result, the frequency was arbitrarily increased by a factor of one hundred over the 1667 Hz resonance frequency of the real Weber bar (see Fig.\ \ref{near-field}).  A coefficient of $GM{L^2}/{c^4}$ was set to unity.  

\section{Riemann tensor for one Weber bar in off-axis direction  }
\label{off}

Interesting Riemann tensor elements are required, for a direction not on a coordinate axis.  Forces on a detector located in the direction 
\begin{equation}
{\bf{n}} = \left( {0.6,0,0.8} \right)
\label{n-num1}
\end{equation}
are evaluated using both the retarded integral and transverse traceless radiation metrics.  For this direction, the harmonic and TT metrics are 
\begin{equation}
{h^{\mu \nu ,H}} = H\left( {\begin{array}{*{20}{c}}
{0.82}&0&0&{0.8}\\
0&{ - 0.18}&0&0\\
0&0&{ - 0.18}&0\\
{0.8}&0&0&{0.82}
\end{array}} \right); \ \ \ \ \ \ 
{h^{\mu \nu ,TT}} = H\left( {\begin{array}{*{20}{c}}
0&0&0&0\\
0&{0.1152}&0&{ - 0.0864}\\
0&0&{ - 0.18}&0\\
0&{ - 0.0864}&0&{0.0648}
\end{array}} \right)
\label{n-num}
\end{equation}
where
\begin{equation}
H =  - \frac{{8GM}}{{{\pi ^2}r{c^4}}}{L^2}\varepsilon {\omega ^2}\cos \left( {\omega t - kr} \right)
\end{equation}
Below is a table of Riemann tensor elements.  In order to evaluate the second derivatives, express the coefficient $H$ as follows
\begin{equation}
H =  - \frac{{8GM}}{{{\pi ^2}r{c^4}}}{L^2}\varepsilon {\omega ^2}\cos \left( {\omega t - k{r_1} - k{\bf{n}} \cdot {\bf{x}}} \right)
\end{equation}
Then
\begin{equation}
{H_{,0}} = \omega H';\ \ \ \ {H_{,1}} =  - \omega {n_x}H';\ \ \ \ {H_{,3}} =  - \omega {n_z}H'
\end{equation}
\begin{equation}
{H_{,00}} = {\omega ^2}H'';\ \ \ \ {H_{,01}} =  - {\omega ^2}{n_x}H'';\ \ \ \ {H_{,03}} =  - {\omega ^2}{n_z}H''
\end{equation}
\begin{equation}
{H_{,11}} = {\omega ^2}n_x^2H'';\ \ \ \ {H_{,13}} = {\omega ^2}{n_x}{n_z}H'';\ \ \ \ {H_{,33}} = {\omega ^2}n_z^2H''
\end{equation}
Augment the table of Riemann tensor elements with numbers, omitting a common factor of ${\omega ^2}H''$.  Different components of the harmonic and TT metrics appear in the Riemann tensor.  However the combinations ``magically'' end up with the same Riemann tensor elements (Table  \ref{tab-riem}).

Incidentally, the gauge transformation (\ref{infinitesimal}) relating the two metrics in eq.\ (\ref{n-num}) is 
\begin{equation}
\left[ {x{'^\mu }} \right] = \left[ {{x^\mu }} \right] - \frac{{8GM}}{{{\pi ^2}r{c^3}}}{L^2}\varepsilon \omega \left( {0.41,0.246,0, - 0.472} \right)\sin \left( {\omega t - kr} \right)
\end{equation}

\begin{table}
\begin{tabular}{ccc}
\hline\noalign{\smallskip}
 & {\bf Harmonic} & {\bf TT}  \\
\noalign{\smallskip}\hline\noalign{\smallskip}
$\bf R_{1010}$ & $ - \frac{1}{2}{g_{00,11}} - \frac{1}{2}{g_{11,00}}=  - 0.0576$ & $ - \frac{1}{2}{g_{11,00}}=  - 0.0576$ \\
 & $[ - 0.5 \cdot 0.82 \cdot {0.6^2} + 0.5 \cdot 0.18] $ &  $[- 0.5 \cdot 0.1152] $ \\
$\bf R_{1030}$ & $ - \frac{1}{2}{g_{00,13}} + \frac{1}{2}{g_{03,10}}=0.0432$ & $ - \frac{1}{2}{g_{31,00}}=0.0432$ \\
$\bf R_{2020}$ & $ - \frac{1}{2}{g_{22,00}}=0.09$ & $ - \frac{1}{2}{g_{22,00}}=0.09$ \\
$\bf R_{3030}$ &\ \ \  $ g_{30,03} - \frac{1}{2}{g_{00,33}} - \frac{1}{2}{g_{33,00}}=0.0324$ & $ - \frac{1}{2}{g_{33,00}}=0.0324$ \\
$\bf R_{3131}$ & $  - \frac{1}{2}{g_{11,33}} - \frac{1}{2}{g_{33,11}}=-0.09$ & $ g_{13,13} - \frac{1}{2}{g_{11,33}}- \frac{1}{2}{g_{33,11}}=-0.09$ \\
$\bf R_{3130}$ & $   \frac{1}{2}{g_{03,13}} - \frac{1}{2}{g_{33,10}}=0.054$ & $   \frac{1}{2}{g_{31,03}}- \frac{1}{2}{g_{33,10}}=0.054$ \\
\noalign{\smallskip}\hline
\end{tabular}
\caption{Elements of the Riemann tensor computed in the direction of eq.\ (\ref{n-num1}). A common factor of ${\omega ^2}H''$ is omitted. The numerical values of the single terms are given explicitly in square brackets only for $ R_{1010}$, as a calculation example.}
\label{tab-riem}
\end{table}

\section{Time-varying fields from point masses  }
\label{time}

An oscillating Weber bar can be considered as a collection of moving point masses.  A summation over point mass pairs produces metric elements consistent with results using a continuous distribution of masses, provided a correct interpretation of the retarded Green’s function is used.  
     Consider point masses located at the top and bottom surfaces of the Weber bar.  The locations of the masses as a function of time are as follows
\begin{equation}
{{\bf{x}}_{Top}}\left( t \right) = \left( {0,0,\frac{1}{2}L + \varepsilon L\cos \left( {\omega t} \right)} \right)
\end{equation}
\begin{equation}
{{\bf{x}}_{Bottom}}\left( t \right) = \left( {0,0, - \frac{1}{2}L - \varepsilon L\cos \left( {\omega t} \right)} \right)
\end{equation}
 The configuration can be described in terms of a conserved stress tensor, 
\begin{equation}
{T^{00}}\left( {{\bf{x}},t} \right) = m{c^2}\delta \left( x \right)\delta \left( y \right)\left[ {\delta \left( {z - {z_{Top}}\left( t \right)} \right) + \delta \left( {z - {z_{Bottom}}\left( t \right)} \right)} \right]
\end{equation}
\begin{equation}
{T^{03}}\left( {{\bf{x}},t} \right) =  - mc\delta \left( x \right)\delta \left( y \right)\left[ {\delta \left( {z - {z_{Top}}\left( t \right)} \right) - \delta \left( {z - {z_{Bottom}}\left( t \right)} \right)} \right]\varepsilon L\omega \sin \omega t
\end{equation}
\begin{equation}
{T^{33}}\left( {{\bf{x}},t} \right) = mL{\omega ^2}\varepsilon \delta \left( x \right)\delta \left( y \right)\theta \left( {{z_{Top}}\left( t \right) - z} \right)\theta \left( {z - {z_{Bottom}}\left( t \right)} \right)\cos \omega t
\end{equation}

The $T^{33}$ element can be interpreted as a massless spring, with a tension ${T_{spring}} = mL{\omega ^2}\varepsilon \cos \omega t$.      The ${\bar h^{00}}$ metric element must be expressed as follows
\begin{equation}
{\bar h^{00}}\left( {{{\bf{x}}_0},{t_0}} \right) = \frac{{4G}}{{{c^2}}}\int {dt'} \int {{d^3}x\frac{{{T^{00}}\left( {{\bf{x}},t'} \right)}}{{\left| {{{\bf{x}}_0} - {\bf{x}}} \right|}}} \delta \left( {{t_0} - t' - \frac{{\left| {{{\bf{x}}_0} - {\bf{x}}} \right|}}{c}} \right)
\end{equation}

The standard technique for computing Lienard-Wiechert potentials requires to perform first the integration over $d^3x$.  Briefly, the $dt'$ integration variable is changed to the resulting argument of the delta function
\begin{equation}
\tau  = t' + \frac{{\left| {{{\bf{x}}_0} - {{\bf{x}}_{Top}}\left( {t'} \right)} \right|}}{c}
\end{equation}
The change in integration variable results in a velocity-dependent factor multiplying the retarded value of the integrand
\begin{equation}
\frac{{dt'}}{{d\tau }} = \frac{1}{{1 - \frac{{{{\hat r}_{Top}} \cdot {{\bf{v}}_{Top}}}}{c}}} =  1 + \frac{{{{\hat r}_{Top}} \cdot {{\bf{v}}_{Top}}}}{c} + {\left( {\frac{{{{\hat r}_{Top}} \cdot {{\bf{v}}_{Top}}}}{c}} \right)^2} + O\left( {\frac{{{v^3}}}{{{c^3}}}} \right)
\label{100}
\end{equation}
where
\begin{equation}
{\hat r_{Top}} = \frac{{{{\bf{x}}_0} - {{\bf{x}}_{Top}}\left( {t'} \right)}}{{\left| {{{\bf{x}}_0} - {{\bf{x}}_{Top}}\left( {t'} \right)} \right|}}; \ \ \ \ \ {{\bf{v}}_{Top}} = \frac{{d{{\bf{x}}_{Top}}\left( {t'} \right)}}{{dt}}
\end{equation}
Similar expressions apply to the point masses on the bottom.  Summing the two contributions 
\begin{equation}
{\bar h^{00}}\left( {{{\bf{x}}_0},{t_0}} \right) = \frac{{4G}}{{{r_0}{c^4}}}{\left[ {2m{c^2} + mc\left( {{{\hat  r}_{Top}} \cdot {{\bf{v}}_{Top}} + {{\hat r}_{Bottom}} \cdot {{\bf{v}}_{Bottom}}} \right)} \right]_{ret}}
\label{103}
\end{equation}
An iterative procedure is used to compute the retarded times, which are slightly different for the two masses.  As a first approximation, 
\begin{equation}
{t_{ret}} = {t_0} - \frac{{\left| {{{\bf{x}}_0}} \right|}}{c} = {t_0} - \frac{{{r_0}}}{c}
\end{equation}
The distances of the two masses from the observation point at the retarded time $t_r$ are 
\begin{equation}
{r_{Top}} = \left| {{{\bf{x}}_0} - {{\bf{x}}_{Top}}\left( {{t_r}} \right)} \right| \cong \left| {{{\bf{x}}_0} - \frac{1}{2}L\hat z} \right| \cong {r_0} - \frac{1}{2}L{n_z}
\end{equation}
\begin{equation}
{r_{Bottom}}  \cong \left| {{{\bf{x}}_0} + \frac{1}{2}L\hat z} \right| \cong {r_0} + \frac{1}{2}L{n_z}
\end{equation}
The velocities of the two masses at the retarded times are computed using an iteration of the retardation times
\begin{equation}
t'_{Top} = {t_0} - \frac{{{r_{Top}}}}{c} = {t_0} - \frac{{{r_0}}}{c} + \frac{{L{n_z}}}{{2c}};\ \ \ \ \ 
t'_{Bottom} = {t_0} - \frac{{{r_{Bottom}}}}{c} = {t_0} - \frac{{{r_0}}}{c} - \frac{{L{n_z}}}{{2c}}
\end{equation}
so that
\begin{equation}
{{\bf{v}}_{Top}} =  - \varepsilon L\omega \sin \left( {\omega {t_0} - k{r_0} + \frac{1}{2}{k_z}L} \right)\hat z;\ \ \ \ 
{{\bf{v}}_{Bottom}} = \varepsilon L\omega \sin \left( {\omega {t_0} - k{r_0} - \frac{1}{2}{k_z}L} \right)\hat z
\end{equation}
For brevity, denote the phase at the origin as $\Phi  = \omega {t_0} - k{r_0}$. The unit vector ${\hat r_{Top}}$ in eq.\ (\ref{103}) is as follows
\begin{equation}
\hat r_{Top} = \frac{{{{\bf{x}}_0} - {{\bf{x}}_{Top}}}}{{\left| {{{\bf{x}}_0} - {{\bf{x}}_{Top}}} \right|}} \cong \frac{{{{\bf{x}}_0} - {{\bf{x}}_{Top}}}}{{{r_0}}}\left( {1 + \frac{{{{\bf{x}}_{Top}} \cdot {\bf{n}}}}{{{r_0}}}} \right) \cong \left( {1 + \frac{{{{\bf{x}}_{Top}} \cdot {\bf{n}}}}{{{r_0}}}} \right){\bf{n}} - \frac{{{{\bf{x}}_{Top}}}}{{{r_0}}} = {\bf{n}} + O\left( {\frac{L}{{{r_0}}}} \right)
\end{equation}
and similarly for ${\hat r_{Bottom}}$.
The distinctions among ${\hat r_{Top}}$, ${\hat r_{Bottom}}$  and ${\bf n}$ are ignored.  

     The time dependent component of ${\bar h^{00}}$ in eq.\ (\ref{103}) involves the sum 
\begin{equation}
\begin{array}{c}
{V_{sum}} = {\bf{n}} \cdot \left( {{{\bf{v}}_{Top}} + {{\bf{v}}_{Bottom}}} \right) = {n_z}\varepsilon \omega L\left( { - \sin \left( {\Phi  + \frac{1}{2}{k_z}L} \right) + \sin \left( {\Phi  - \frac{1}{2}{k_z}L} \right)} \right) = \\
=  - 2{n_z}\varepsilon L\omega \cos \Phi \sin \left( {\frac{1}{2}{k_z}L} \right) \cong  - n_z^2\varepsilon \frac{{{\omega ^2}}}{c}{L^2}\cos \Phi 
\end{array}
\end{equation}
Hence
\begin{equation}
{\bar h^{00}}\left( {{{\bf{x}}_0},{t_0}} \right) = \frac{{4G}}{{{r_0}{c^4}}}\left[ {2m{c^2} - m{\omega ^2}{L^2}n_z^2 \varepsilon \cos \Phi } \right]
\end{equation}
The retarded integrals of $T^{03}$ and $T^{33}$ can now be computed, ignoring the velocity-dependent factor of eq. (\ref{100})
\begin{equation}
{h^{03}}\left( {{{\bf{x}}_0},{t_0}} \right) = \frac{{4G}}{{{r_0}{c^4}}}\left( { - mc\varepsilon L\omega } \right)\left[ {\sin \left( {\Phi  + \frac{1}{2}{k_z}L} \right) - \sin \left( {\Phi  - \frac{1}{2}{k_z}L} \right)} \right] =  - \frac{{4G}}{{{r_0}{c^4}}}mc\varepsilon {L^2}{\omega ^2}{n_z}\cos \Phi 
\end{equation}
\begin{equation}
{\bar h^{33}}\left( {{{\bf{x}}_0},{t_0}} \right) =  - \frac{{4G}}{{{r_0}{c^4}}}\left( {mc\varepsilon L{\omega ^2}} \right)L\cos \Phi 
\end{equation}
Hence
\begin{equation}
{\bar h^{\mu \nu }}\left( {{{\bf{x}}_0},{t_0}} \right) =  - \frac{{4G}}{{{r_0}{c^4}}}m{\omega ^2}\varepsilon{L^2}\left( {\begin{array}{*{20}{c}}
{n_z^2}&0&0&{{n_z}}\\
0&0&0&0\\
0&0&0&0\\
{{n_z}}&0&0&1
\end{array}} \right) \cos \left( {\omega {t_0} - k{r_0}} \right)
\end{equation}

\subsection{Sum over mass pairs  }

      The general relation between the location $Z_i$ of a point mass and its original position $z_i$ is 
\begin{equation}
 {Z_i} = {z_i} + w = {z_i} + \varepsilon L\sin \left( {\frac{{\pi {z_i}}}{L}} \right)\cos \omega t
\end{equation}
Sum (integrate) $Z^2$ over all mass pairs, 
\begin{equation}
 Z_i^2 = {\left( {{z_i} + \varepsilon L\sin \left( {\frac{{\pi {z_i}}}{L}} \right)\cos \omega t} \right)^2} \cong z_i^2 + 2{z_i}\varepsilon L\sin \left( {\frac{{\pi {z_i}}}{L}} \right)\cos \omega t
\end{equation}
The time dependent part of the second moment is 
\begin{equation}
 {Q_{zz}}\left( t \right) = \sum\limits_{pairs} {2{m_i}} {z_i}\varepsilon L\sin \left( {\frac{{\pi {z_i}}}{L}} \right)\cos \omega t
\label{120}
\end{equation}
Replace $m_i$ by ${m_i} = M \Delta {z_i}/{L}$, to convert the sum (\ref{120}) to an integral 
\begin{equation}
 {Q_{zz}}\left( t \right) = \frac{M}{L}\int\limits_{ - L/2}^{L/2} {dz} 2z\varepsilon L\sin \left( {\frac{{\pi z}}{L}} \right)\cos \omega t
= 4\varepsilon M\frac{{{L^2}}}{{{\pi ^2}}} \cos \omega t
\end{equation}
     The sum over mass pairs gives the result 
\begin{equation}
 {\bar h^{\mu \nu }}\left( {{\bf{x}},t} \right) =  - \frac{{ - 8}}{{{\pi ^2}}}\frac{G}{{r{c^4}}}M{L^2}\varepsilon {\omega ^2}
\left( {\begin{array}{*{20}{c}}
{n_z^2}&0&0&{{n_z}}\\
0&0&0&0\\
0&0&0&0\\
{{n_z}}&0&0&1
\end{array}} \right)\cos \left( {\omega t - kr} \right)
\end{equation}
The result agrees with that computed using density as a continuous distribution.

\section{Further possible applications}
\label{app}

As we mentioned in the Introduction, the computational scheme presented in this paper can be applied to physical sources for which the usual quadrupolar formula is not adequate.

Examples of sources which do not have a symmetry compatible with the TT gauge are the arrays of mechanical or electromagnetic oscillators devised for the laboratory generation of high-frequency gravitational waves (HFGWs). Such generators have been discussed, among others, by Grishchuk \cite{Grischchuk1}. They are still far from concrete realization, but it has been estimated that they could generate HFGWs with amplitudes only 100 times smaller than the main astrophysical sources. 

{\it Retardation effects in the source} occur when the wavelength of the gravitational wave is comparable to the size of the source, so that the waves generated at different points do not add in phase. 
A typical example are electromagnetic sources made of resonant cavities (see \cite{Grischchuk2} for an exact solution with a thoroidal cavity). One can also conceive an intermediate case, i.e.\ generators which contain stationary waves of a material medium, having propagation velocity much larger than the sound velocity of compression waves in Weber bars. Also in this case the retardation effects are important.

Consider for instance a plasma  supporting stationary waves of the Alfven type. These waves can have propagation velocity close to $c$ and frequency up to several MHz. Their wavelength is then such to produce retardation effects in the source. It is interesting to observe that in this case the emitted gravitational wave can have an oscillating-dipole component without violating the conservation of momentum in the source.

\begin{figure}
\begin{center}
  \includegraphics[width=10cm,height=7cm]{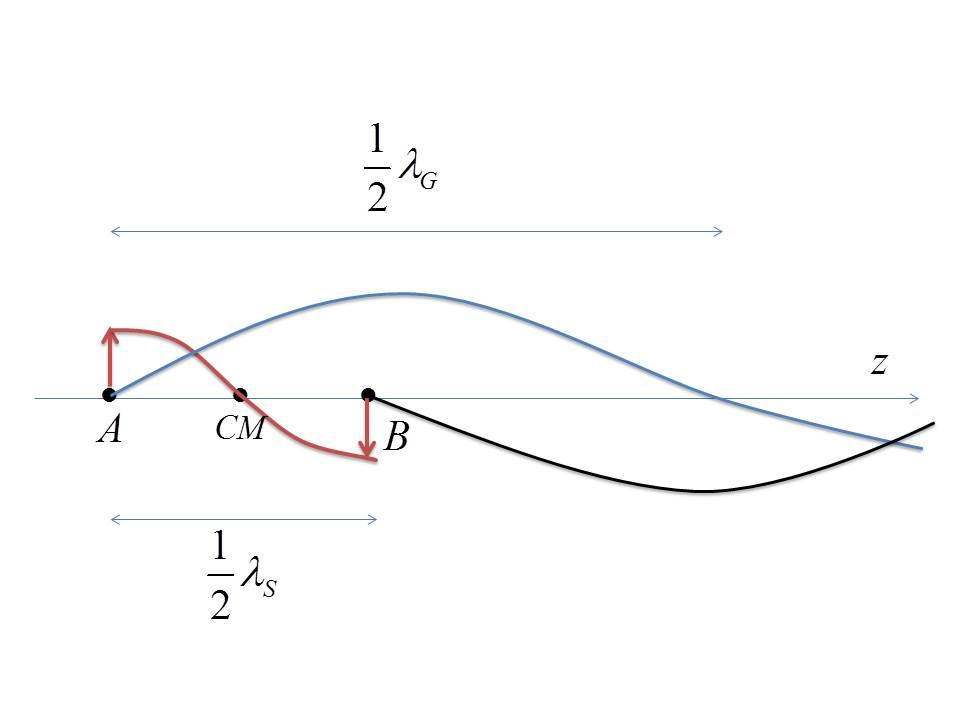}
\caption{Elementary illustration of retardation effects in a source of gravitational waves (plasma with high-speed transversal stationary waves). Two mass elements $A$ and $B$ of the plasma oscillate in phase opposition (red), so that their center of mass is at rest. The dipolar waves generated by $A$ and $B$ (blue and black) would cancel if their wavelength $\lambda_G$ was much larger  than the length $\lambda_S$ of the stationary ``matter wave'' (as it usually happens). In this case, however, the two wavelengths are comparable and the negative interference (here only shown in the $z$ direction) is only partial. Note that in reality the near field is different from a pure sinusoidal function; the figure only shows that the argument for the total negative interference of the dipolar components is not applicable.
}
\label{plasma-wave} 
\end{center}
   \end{figure}

Namely, consider in the stationary plasma wave of wavelength $\lambda_S$ (red in Fig.\ \ref{plasma-wave}) two points $A$ and $B$, which are opposite with respect to the center of mass of the plasma. The plasma particles at these points move in opposite directions (red arrows), thus keeping the center of mass at rest. Each of these elementary oscillators generates dipolar waves with opposite phases (black and blue lines in Fig. \ref{plasma-wave}) and wavelength $\lambda_G$. Such waves would interfere destructively if their wavelength was much larger than the distance between $A$ and $B$; but in this case the interference is not entirely destructive. 

Note that the ``dipolar component'' generated in this way has an amplitude comparable to the quadrupolar component, because in general the amplitude ratio of the two components is of the order of $d/\lambda_G$, where $d$ is the size of the source. In fact, the multipolar expansion should be replaced by the computation of a full retarded integral.

\section{Conclusions}

    This document uses ``pedestrian'' but comprenhensive techniques to show a connection between the Lorentz-harmonic (retarded integral) and TT (transverse traceless) metrics.  The local and global aspects of this connection are emphasized. The harmonic metric includes a propagating gravitoelectric field, which is absent in the TT metric.  Gradients in the gravitoelectric field combine with tidal forces from velocity dependent terms in the geodesic equation, resulting in purely transverse local disturbances in the far field.  
     The concepts of gravitoelectric and gravitomagnetic fields are valid in regions wherein time derivatives of spatial components of the metric are negligible.  A cross-over from near-field geodesic forces and far-field tidal forces occurs at a distance of a wavelength from a time-varying source.

The physical effects of the metrics are evaluated for a general kind of detector composed of two parallel rings, which are displaced and deformed by the field. The harmonic metric is computed also in the near-field region, both analytically and numerically. The analogues of the displacement current and Faraday induction are displayed. A suitable application of the retarded Green function allows to re-obtain the harmonic metric of the oscillating Weber bar as an integral over point masses. Some possible further applications of this technique are discussed, e.g.\ to the field generated by a stationary wave in a plasma, which is expected to exhibit relevant effects of retardation in the source.



\end{document}